\documentclass[prb,a4paper,twocolumn,showpacs,floatfix,superscriptaddress]{revtex4}
\usepackage{graphicx}
\usepackage{amsmath,amssymb}
\usepackage{color}

\begin{document}

\title[Transition to complete synchronization in phase coupled oscillators]{Transition to complete synchronization in phase coupled oscillators with nearest neighbours coupling}

\author{Hassan F. El-Nashar}

\affiliation{Department of Physics, Faculty of Science, Ain Shams University,
11566 Cairo, Egypt}

\affiliation{Department of Physics, Faculty of Education, King Saud University, P.O.~Box 21034, 11942 Alkharj, K.S.A}

\author{Paulsamy Muruganandam}

\affiliation{School of Physics, Bharathidasan University, Palkalaiperur,
Tiruchirappalli -- 620024, India}

\author{Fernando F. Ferreira}

\affiliation{Grupo Interdisciplinar de F\'{\i}sica da Informa\c{c}\~ao e Economi a (GRIFE), Escola de Arte, Ci\^encias e Humanidades, Universidade de S\~ao Paulo, Av. Arlindo Bettio 1000, 03828-000 S\~ao Paulo, Brazil}

\author{Hilda A. Cerdeira}

\affiliation{Instituto de F\'{\i}sica Te\'orica, Universidade Estadual
Paulista, R.  Pamplona 145, 01405-000 S\~ao Paulo, Brazil}

\affiliation{Instituto de F\'{\i}sica, Universidade de S\~ao Paulo, R.
do Mat\~ao, Travessa R. 187, 05508-090 S\~ao Paulo, Brazil}


\begin{abstract}

We investigate synchronization in a Kuramoto-like model with nearest neighbour coupling. Upon  analyzing the behaviour of individual oscillators at the onset of complete synchronization, we show that the time interval between bursts in the time dependence of the frequencies of the oscillators exhibits universal scaling and blows up at the critical coupling strength. We also bring out a key mechanism that leads to phase locking. Finally, we deduce forms for the phases and frequencies at the onset of complete synchronization.

\end{abstract}



\maketitle

{\bfseries Weakly coupled oscillators play an important role in understanding collective behaviour of large populations. They are often used to model the dynamics of a variety of systems that arise in nature, even though they are quite different. Synchronization is one of the interesting phenomena observed in these systems where the interacting oscillators under the influence of coupling would have a common frequency. Particularly, these systems show an extremely complex clustering behavior as a function of the coupling strength. In spite of their differences, the above mentioned systems can be described using simple models of coupled phase equations such as the Kuramoto model. This paper analyzes the behaviour of individual oscillators in the vicinity of the critical coupling where all the oscillators evolve in synchrony with each other.}

\section{Introduction}
Systems of coupled oscillators can describe problems in physics, chemistry, biology, neuroscience and other disciplines. They have been widely used to model several phenomena such as: Josephson
junction arrays, multimode lasers, vortex dynamics in fluids, biological information processes, neurodynamics \cite{1,2,14}. These systems have been observed to synchronize themselves to a
common frequency, when the coupling strength between these oscillators is increased \cite{5,6,10,13,15,7,8,9,11,is}. The synchronization features of many of the above mentioned systems, in spite of the diversity of the dynamics, might be described using simple models of weakly coupled phase oscillators such as the Kuramoto model \cite{15,17}.

Finite range interactions are more realistic for the description of many physical systems, although finite range coupled systems are difficult to analyze and to solve analytically. However, in order to figure out the collective phenomena when finite range interactions are considered, it is of fundamental importance to study and to understand the nearest neighbour interactions, which is the simplest form of the local interactions. In this context, a simplified version of the Kuramoto model with nearest neighbour coupling in a ring topology, which we shall refer to as \emph{locally coupled Kuramoto model} (LCKM), represents a good candidate to describe the dynamics of coupled systems with local interactions such as Josephson junctions, coupled lasers, neurons, chains with disorders, multi-cellular systems in biology and in communication systems \cite{10,17,arx,wie1}. For instance, it has been shown that the equations of the resistively shunted junction which describe a ladder array of overdamped, critical-current disordered Josephson junctions that are current biased along the rungs of the ladder can be expressed by a LCKM  \cite{16}. In nearest neighbours coupled R\"ossler oscillators the phase synchronization can be described by the LCKM \cite{22}. Therefore, LCKM can provide a way to understand phase synchronization in coupled systems, for example, in locally coupled lasers \cite{wie2,wie3}, where local interactions are dominant. Coupled phase oscillators described by LCKM can also be used to model the occurrence of travelling waves in neurons \cite{10,suz}. In communication
systems, unidirectionally coupled Kuramoto model can be used to describe an antenna array~\cite{iop}. Such unidirectionally coupled Kuramoto models can be considered as a special case of the LCKM and it often mimics the same behaviour.

One of the important features of the local model is that the properties of individual oscillators can be easily analyzed to study the collective dynamics while one has to rely on the average quantities, in a mean field approximation or by means of an order parameter, etc., as in the
case of the usual Kuramoto model of long range interactions. Therefore, due to the difficulty in applying standard techniques of statistical mechanics, one should look for a simple approach to understand the coupled system with local interactions by means of numerical study of a temporal behaviour of the individual oscillators. Such analysis is necessary in order to get a close picture on the effect of the local interactions on synchronization. In this case, numerical investigations can assist to figure out the mechanism of interactions at the stage of
complete synchronization which in turns help to get an analytic solution. Earlier studies on the LCKM show several interesting features including tree structures with synchronized clusters, phase slips, bursting behaviour and saddle node bifurcation and so on \cite{Strogatz1988,Zheng98,Zheng2000}. There have been studies showing that neighbouring elements share dominating frequencies in their time spectra, and that this feature plays an important role in the dynamics of formation of clusters in the local model \cite{19,m}. It has been found that the order parameter, which measures the evolution of the phases of the nearest neighbour oscillators, becomes maximum at the partial synchronization points inside the tree of synchronization \cite{20}. Very recently we developed a scheme based on the method of Lagrange multipliers to estimate the critical coupling strength for complete synchronization in the local Kuramoto model with different boundary conditions \cite{21}.

In this paper we address the mechanism that leads to a complete synchronization in the Kuramoto model with local coupling. This is done by analyzing the behaviour of each individual oscillators at the onset of synchronization. For this purpose we consider the equations governing the phase differences at the onset of synchronization. In particular, we identify that the cosine of
only one among the phase differences becomes zero. Based on this property we derive the expression for the time interval between bursting behaviour of the instantaneous frequencies of each individual oscillators in the vicinity of critical coupling strength. Our analysis shows that the transition to complete synchronization occurs due to a saddle node bifurcation in
agreement with the earlier studies. Further we deduce the expressions for the phases and frequencies of the individual oscillators at the onset of complete synchronization.

This paper is organized as follows. In Sec.~\ref{mechanism} we
present a brief overview on the dynamics of local Kuramoto model.
Then we analyze the behaviour of the phase differences and the
time interval between successive bursts at the transition to
complete synchronization. In particular, we point out the
mechanism that lead to complete phase locking at the critical
coupling strength. Based on this we deduce the forms of phases and
frequencies at the onset of synchronization. Finally, in
Sec.~\ref{finalwords} we give a summary of the results and
conclusions.

\section{Behaviour of phases and frequencies at the onset of
synchronization}
\label{mechanism}

Even when there has been an extensive exploration of the dynamics
of the Kuramoto model (global coupling among all oscillators), the
local model of nearest neighbour interactions, which can be
considered as a diffusive version of the Kuramoto model,
has been receiving attention only recently. The LCKM is expressed
as \cite{Zheng2000,19,m,20,21}:
\begin{align}
\dot{\theta}_i=\omega_i+\frac{K}{3}[\sin(\theta_{i+1}-\theta_i) +
\sin(\theta_{i-1} - \theta_i)], \label{kura:theta}
\end{align}
here $\omega_i$ are the natural frequencies, $K$ is the coupling
strength, $\theta_i$ is the instantaneous phase, $\dot{\theta_i}$
is the instantaneous frequency and $i=1,2,\ldots,N$. Many
interesting features of the LCKM remain unknown, especially an
analytic solution \cite{17}, which would be of great importance in
understanding the mechanism that leads to synchronization. In
order to find such an analytic solution, one should study
carefully the temporal evolution of frequency and phase of each
individual oscillator in the neighbourhood of the critical
coupling for complete synchronization.

If we consider the oscillators in a ring, with periodic boundary
conditions $\theta_{i+N}=\theta_i$, the nonidentical oscillators
(\ref{kura:theta}) cluster in time averaged frequency, until they
completely synchronize to a common value of the average frequency
$\omega_0 = \frac{1}{N}\sum_{i=1}^{N}\omega_i$, at a critical
coupling $K_c$ \cite{Zheng98,Zheng2000,19,20,21}.
\begin{figure}[!ht]
\centering\includegraphics[width=\linewidth,clip]{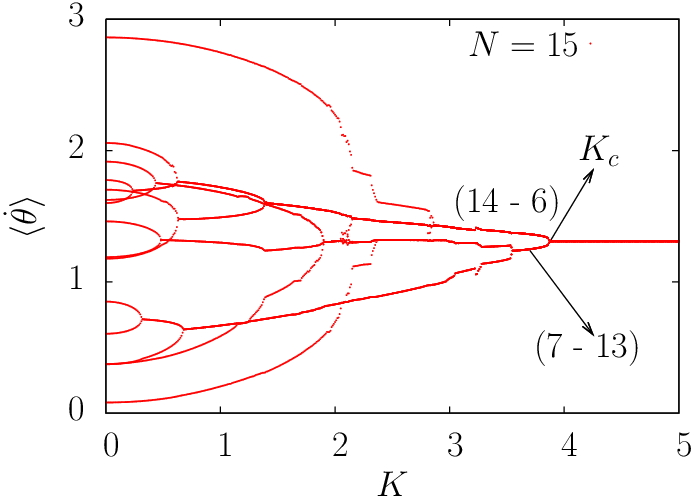}
\caption{(Color online) Synchronization tree for a system of $15$ oscillators.}
\label{fig:kura_tree}
\end{figure}
At $K \ge K_c$  the phases and the frequencies are time
independent and all the oscillators remain synchronized. In
Fig.~\ref{fig:kura_tree}, we show the synchronization tree for a
periodic system with $N=15$ oscillators, where the elements which
compose each one of the major clusters that merge into one at
$K_c$, are indicated in each branch.

In terms of phase differences $\phi_i = \theta_{i+1} - \theta_{i}$, system
(\ref{kura:theta}), can be rewritten as:
\begin{align}
\dot \phi_i = \omega_{i+1} - \omega_{i} + \frac{K}{3} \left[ \sin
\phi_{i-1}  - 2 \sin \phi_{i} + \sin \phi_{i+1} \right],
\label{kura:phase_diff}
\end{align}
with $\dot \phi_i^*=0$ at $K_c$ for $i=1,2,\ldots, N$. In
addition, all quantities $\dot\phi_i^*$, $\phi_i^*$ and $\dot
\theta_i^*$, which become time independent \cite{Zheng2000,19,20,21} at
the critical coupling, remain like that $k \geq K_c$ when
$\dot\phi_i^*=0$ and $\dot \theta_i^*= \omega_0$. Earlier
attempts to get a solution of the above
eq.~(\ref{kura:phase_diff}) show that for only two oscillators
which have phase difference $\phi_l^*=\theta_l^* -
\theta_{l-1}^*$, results $\vert \sin\phi_l^*\vert =1$ \cite{Strogatz1988} and indeed this is
a necessary condition for eq.~(\ref{kura:phase_diff}) to have a
phase-locked solution. This fact has been used by Daniels et al
\cite{16} to estimate the value of critical coupling strength
$K_c$, at which the transition to complete synchronization
occurs. However the determination of which two oscillators among
$N$ oscillators that have
$\vert\sin\phi_l^*\vert=1$, remains difficult. From the study of
the temporal evolution of phases and frequencies of each
individual oscillator, it has also been found numerically that,
at the onset of synchronization $K \lesssim K_c$, the values of
$\dot \theta_i(t)$ and $\dot \phi_i(t)$ remain equal to
$\omega_0$ and zero, respectively, for a certain time interval
$T$. During this time $T$ a stable phase-locked solution exists,
then they burst \cite{Zheng98,Zheng2000,22}, and this stable phase-locked
solution is lost. In between bursts, the phases remain fixed and
then they have an abrupt change (phase-slip behaviour) by an
amount which depends on the initial values of the frequencies
${\omega_i}$ \cite{Zheng98,Zheng2000}, corresponding to the burst in the
frequencies, while the quantities $\sum_{i=1}^{N}\phi_i=0$ and
$\sum_{i=1}^{N}\dot \phi_i=0$ are always preserved by the
topology. Integrated with the above information, it has been
shown by numerical investigation that the time interval $T$ blows
up as $K$ becomes close to $K_c$ and $T  \to \infty$ at $K_c$.
All these information leads to conclude that there is a
saddle-node bifurcation at $K_c$ and the
synchronization-desynchronization transition at the critical
coupling can be interpreted using this knowledge.

In this work, we perform numerical investigations of the temporal
evolution of the phases and frequencies for the individual
oscillators in order to arrive to specific conditions which will
lead to criteria to obtain an analytic solution. A detailed study
of all quantities $\sin\phi_i^*$ at $K_c$ for several values of
$N$ and for different sets of $\omega_i$, shows that there is
only one value of phase difference between two neighbouring
oscillators $\phi_l^* = \theta_{l+1}^* - \theta_l^*$ for which
$\vert\sin\phi_l^*\vert=1$, while for all other values, $i \neq
l$ $\vert\sin\phi_i^*\vert \neq 1$.
\begin{figure}[!ht]
\centering\includegraphics[width=\linewidth,clip]{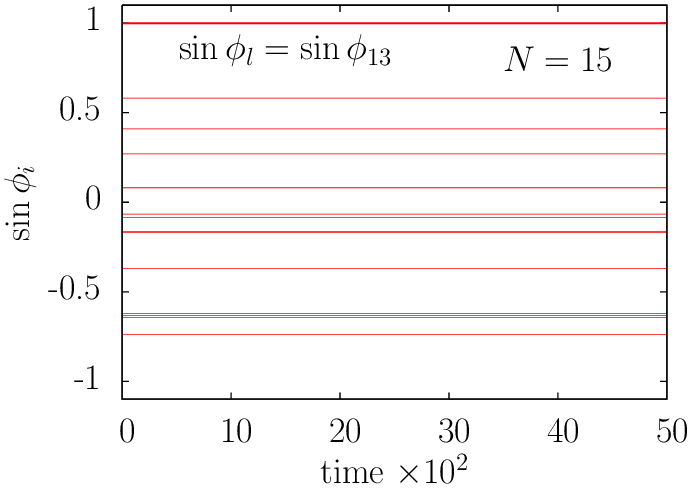}
\caption{(Color online) Values of $\sin \phi_i$ at $K \ge K_c$ for a system of
$15$ oscillators [see Fig.~\ref{fig:kura_tree}].}
\label{kura_sins}
\end{figure}
In Fig.~\ref{kura_sins}, we show $\sin\phi_i^*$ for a case of
$N=15$ as time progresses at the critical coupling $K_c$, with
the same initial frequencies of Fig. 1. We see that the value of
$\vert\sin\phi_l^*\vert = 1$, is for $l=13$ and that this
quantity $\vert\sin\phi_l^*\vert=1$ holds for only one value of
phase difference $\phi_l=\pi/2$ where these two oscillators $l+1$
and $l$ belong to different clusters, and these two nearest
neighbours oscillators are always at the borders between the
major clusters that merge at $K_c$, which can be seen from Fig.
1. We find the same result for different initial frequencies
$\omega_i$ and for different values of $N$. In addition, the sign
of $\sin\phi_l^*$ is negative for $\omega_l > \omega_{l+1}$ and
positive for the reverse.

\begin{figure}[!ht]
\centering\includegraphics[width=\linewidth,clip]{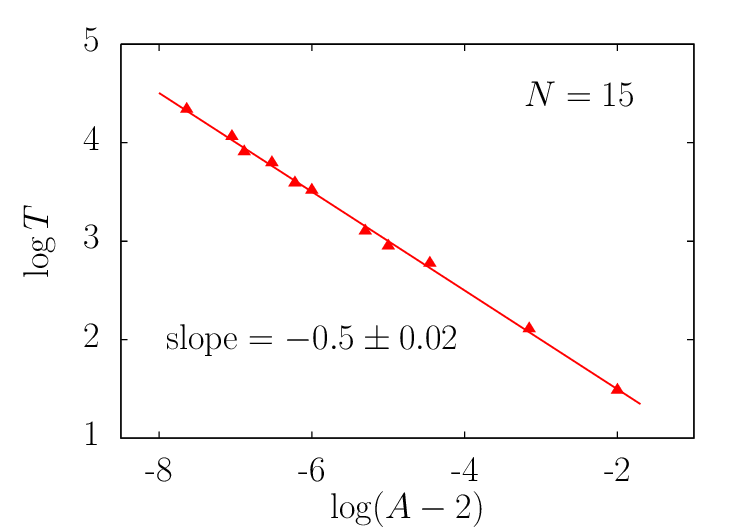}
\caption{(Color online) $\log T$ versus $\log \left(A-2\right)$, which shows the divergence of the time interval $T$ when $A$ approaches $2$ with
a slope $\simeq -0.5$ .} \label{kura_A}
\end{figure}

The knowledge of the burst and phase slip (in the vicinity of
$K_c$) of the quantities $\dot \phi_i(t)$ and $\phi_i(t)$,
respectively, as well as the finding of
$\vert\sin\phi_l^*\vert=1$ (at $K_c$), will allow us to rewrite
equation (\ref{kura:phase_diff}), for the index $l$ as:
\begin{align}
\dot \phi_l= B \left( A-2\sin\phi_l \right), \label{non1}
\end{align}
where $A=\frac{3|(\omega_{l+1}-\omega_l)|}{K} + \sin\phi_{l-1} +
\sin\phi_{l+1}$ and $B=\frac{K}{3}$. Eq.~(\ref{non1}) takes the
form of a phase synchronization of two coupled limit-cycles
\cite{stro}. At $K_c$, $\phi_l^*=\pm \pi/2$ and, $\phi_{l-1}^*$
and $\phi_{l+1}^*$ are constants and time independent and $A=2$.
A detailed numerical study shows also that, at the onset of
synchronization, $A\approx2$ and the values of $\phi_l$,
$\phi_{l-1}$ and $\phi_{l+1}$ remain equal to their values at
$K_c$, for a time interval $T$. The values of $K_c$ and $A$ for
different number of oscillators $N$ from numerical simulations
are tabulated in Table.~\ref{avalues}. It is clear
\begin{table}[!ht]
\begin{center}
\caption{Calculated values of $K_c$ and $A$ for different values of
$N$.} \label{avalues}
\begin{tabular}{crr|crr}
\hline \multicolumn{1}{c}{~~N~~}
    & \multicolumn{1}{c }{$K_c$} & \multicolumn{1}{c }{$A$} &
\multicolumn{1}{|c}{~~N~~}
    & \multicolumn{1}{c }{$K_c$} & \multicolumn{1}{c }{$A$} \\
\hline
  3 &  0.85041227~  & 1.9994~  &  20 &  4.95830014~  & 2.0002~   \\
  5 &  3.17082713~  & 2.0001~  &  25 &  3.64106038~  & 1.9989~   \\
 10 &  3.54701035~  & 1.9996~  &  50 &  9.45720049~  & 1.9993~   \\
 15 &  3.87023866~  & 2.0000~  & 100 &  12.7232087~  & 1.9985~   \\
\hline
\end{tabular}
\end{center}
\end{table}
that in all the cases, $A\simeq 2$ when $K$ approaches $K_c$. The
relation $A\simeq 2$ is found to be valid for different choices
of initial frequencies $\omega_i$ for each $N$ in the vicinity of
$K_c$. Further, it should be noted that when the time interval $T
\rightarrow \infty$ one can find that $A=2$. The time interval
$T$ can be found analytically, according to eq.~(3), to be
\begin{align}
T \approx \frac{3 \pi \sqrt{2}}{K \sqrt{A}}\frac{1}{\sqrt{A-2}}.
\label{eq:T1}
\end{align}
In Fig.~\ref{kura_A} we clearly see that $T$ blows up as $A$
becomes close to $2$ (where $K$ goes to $K_c$), for the case of
$N=15$. We find that $T$
blows up as $(A-2)^{-0.5}$ which is a numerical proof that a saddle-node
bifurcation occurs at $K_c$. Assuming that $\sin\phi_{l-1}$ and
$\sin\phi_{l+1}$ remain constant for a time interval $T$, in the
vicinity of $K_c$, and equal to their values at $K_c$ (which has
been verified numerically), we find that $\frac{AK}{2} \approx
K_c$.
\begin{table}[!ht]
\begin{center}
\caption{Calculated values of $\frac{AK}{2}$ for $N=15$
oscillators at the vicinity of $K_c=3.870238658$.}
\label{avalues1}
\begin{tabular}{c|c|c|c}
\hline
  K           &  A      &$\frac{AK}{2}$&$|K_c-K|$ \\[1mm] \hline
  3.869480136~ & ~2.0007000~ & ~3.870834454~ & ~5.960$\times 10^{-4}$~ \\
  3.870198414~ & ~2.0000350~ & ~3.870266142~ & ~2.750$\times 10^{-5}$~ \\
  3.870226709~ & ~2.0000100~ & ~3.870246060~ & ~7.402$\times 10^{-6}$~ \\
  3.870237122~ & ~2.0000010~ & ~3.870238909~ & ~2.480$\times 10^{-7}$~ \\
  3.870238325~ & ~2.0000003~ & ~3.870238727~ & ~6.920$\times 10^{-8}$~ \\ \hline
\end{tabular}
\end{center}
\end{table}
Table~\ref{avalues1} shows this fact where the error is small and
decreases as $K$ approaches $K_c$. Therefore, eq. ~(\ref{eq:T1})
takes the form
\begin{align}
T\approx\frac{3 \pi }{\sqrt{2}\sqrt{K_c}\sqrt{K_c-K}}.
\label{eq:T2}
\end{align}
So that, within a good approximation, the periodic time interval
$T$ blows up as $(K_c-K)^{-0.5}$, in good agreement with the
numerical calculation by Zheng et. al \cite{Zheng98,Zheng2000}, showing
that a saddle-node bifurcation occurs at $K_c$~\cite{Strogatz1988}.

Therefore, eq.~(\ref{non1}) can be written as
\begin{align}
\dot \phi_l \approx L\left( K_c-K\sin\phi_l \right), \label{non2}
\end{align}
which can be solved analytically and its solution reads
\begin{subequations}
\label{eq7}
\begin{align}
\phi_l \approx  2\arctan\left [\frac{\alpha \tan
\left(\frac{1}{2}\alpha Lt\right) \pm K}{K_c}\right],
\label{eq7a}
\end{align}
and
\begin{align}
\dot \phi_l \approx \frac{L}{K_c}
\frac{\alpha^2 \sec^2\left(\frac{1}{2}\alpha Lt\right)}{1+
\left\{\frac{1}{K_c}\left[
\alpha \tan \left(\frac{1}{2}\alpha Lt\right)\pm K\right]\right\}^2}, \label{eq7b}
\end{align}
\end{subequations}
where $\alpha = \sqrt{K_c^2-K^2}$ and $L=\frac{2}{3}$.
Eqs.~(\ref{eq7a}) and ~(\ref{eq7b}) show that, at $K_c$, the
values $\sin\phi_l^*= \pm 1$ which lead to $\dot \phi_l^*=0$.
Also, it can be seen that in the vicinity of $K_c$, $\sin\phi_l=
\pm 1$ and $\dot \phi_l=0$ for a period $T$. The $(+)$ sign
in Eqs. (7) is corresponding to the case
$\omega_{l+1}>\omega_l$ and the $(-)$ sign for the reverse.

In order to understand the mechanism of full synchronization
which occurs at $K_c$, we use the fact that $\sin\phi_l^*=\pm 1$
and each $\dot \theta_i^*=\omega_0$, where these quantities
remain unchanged for $T$ in the vicinity of $K_c$. Hence, from
system (1), we are able to get the following relations:
\begin{subequations}
\begin{align}
\sin\phi_{l+m}^*= &\, \frac{3}{K_c}
\sum_{m=1}^{N-l}(\omega_0-\omega_{l+m}) \pm \sin\phi_l^*, \\
\sin\phi_{l-n}^*= &\,
-\frac{3}{K_c}\sum_{n=1}^{l-1}(\omega_0-\omega_{l-n-1}) \pm
\sin\phi_l^*.
\end{align}
\end{subequations}
Using this fact, we write the following equations, in addition to
eq.~(\ref{eq7}):
\begin{subequations}
\begin{align}
\phi_{l-n} \approx &\, \sin^{-1}(a_n \pm \sin\phi_l), \\
\dot \phi_{l-n} \approx &\, \frac{\cos\phi_l \, \dot\phi_l}{\sqrt{1-(a_n \pm \sin\phi_l)^2}}, \\
\phi_{l+m} \approx &\, \sin^{-1}(a_m \pm \sin\phi_l),\\
\dot \phi_{l+m} \approx &\, \frac{\cos\phi_l \, \dot
\phi_l}{\sqrt{1-(a_m \pm \sin\phi_l)^2}},
\end{align}
\end{subequations}
where $a_n=\frac{-3}{K_c} \sum_{i=1}^n(\omega_0-\omega_{l-i-1})$
with $n=1,2,3,\ldots,l-1$ and $a_m=\frac{3}{K_c}
\sum_{j=1}^m(\omega_0-\omega_{l+j})$ with $m=1,2,3, \ldots, N-l$. It
is clearly seen that according to the above equation, each
$\phi_i$ can be expressed in terms of $\phi_l$ and consequently
\begin{figure}[!ht]
\centering\includegraphics[width=\linewidth,clip]{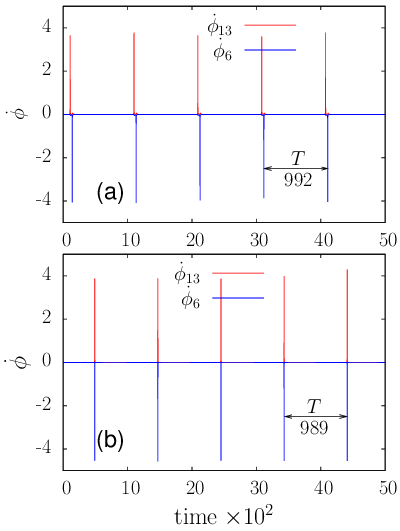}
\caption{(Color online) Time evolution of $\dot\phi_{13}$ and $\dot\phi_6$
according to (a) system (1) and (b) equations (7b) and (9b), at
$K=3.870226709$, for $15$ oscillators with the same initial
conditions of Fig. 1} \label{fig_compare}
\end{figure}
each $\dot \phi_i$ can be expressed in terms of $\phi_l$ and $\dot
\phi_l$. Therefore, all values of $\phi_i$ will be shifted from
each other by some constant which is determined by the location of
the indexes $l-n$ and $l+m$ relative to oscillators with indexes
$l$ and $l+1$. This is shown in Fig. (2), where $\sin\phi_i$
values are shifted from each others at $K_c$. Therefore, at $K_c$,
what occurs to $\phi_l$ and $\dot \phi_l$ due to saddle-node
bifurcation diffuses through the ring via interaction between
neighbouring oscillators. This means that, at the vicinity of
$K_c$, the value of $\phi_l$ has an abrupt change after being
constant for a time $T$, caused by a burst behaviour of $\dot
\phi_l$ after being zero for the same time interval $T$. The
abrupt change in $\phi_l$ produces a sudden change in the values
of $\phi_i$ of their neighbours, while the bursting behaviour of
$\dot \phi_l$ in turn yields bursts in $\dot \phi_i$ ($i \neq l$).
In order to demonstrate this fact, we plot the temporal evolution
of $\dot\phi_{13}$ and $\dot\phi_6$, in the vicinity of $K_c$,
according to numerical simulation of eq. (1) in Fig. 4a while we
plot both quantities according to eq. (7b) and eq. (9b) in Fig.
4b. As shown in Fig. 4, the results of numerical simulation agrees
with that from the analytic solution. The above mentioned
behaviour is reflected in the time dependence of the $\dot
\theta_i's$, which in turn remain equal to $\omega_0$ for a time
$T$ and burst around $\omega_0$ corresponding to the burst of
$\phi_l$. Henceforward, we argue that it is the behaviour of
$\phi_l$ and $\dot \phi_l$, which drives the system to fall into
full synchronization.

\section{Summary and conclusions}
\label{finalwords}

In summary, we have analyzed the conditions on the phase differences for the
onset of complete synchronization at the critical coupling strength in a
Kuramoto-like model with nearest neighbour coupling. Such condition, which
is $\vert \sin\phi_l^*\vert =1$ (or $\cos\phi_l^* = 0$), allows us to solve
the equations of the phase differences (\ref{kura:phase_diff}) analytically.
Also, we found that full synchronization occurs always when the quantity $A=2$
at $K_c$. Due to the diffusive nature of the LCKM, complete synchronization of
all oscillators to a common value can be interpreted and understood once we
have an analytic forms for $\phi_l$ and $\dot \phi_l$. However, it is still
difficult to determine analytically the number of oscillators in each cluster
which merge into one at $K_c$. Therefore, one cannot allocate straightforwardly
the two nearest neighbour oscillators which would have $\vert\sin\phi_l^*\vert
= 1$. On the other hand, a detailed numerical study on the temporal evolution
of phases, phase-differences and frequencies of oscillators at the borders of
the clusters that merge into larger one at onset of complete synchronization
helps us to determine the neighbouring oscillators which have $\sin\phi_l^* =
\pm 1$. Such analysis can also be used to understand the partial
synchronization that leads to the formation of small clusters for coupling
strengths below the critical coupling strength $K_c$.
Of course analysis of the simplest case of
locally coupled phase oscillators can help to understand models with local
interactions where amplitudes and phases are included \cite{arx,22,wie2,wie3}. In such cases, a detailed
study of the time evolution of amplitudes and phases can reveal a better
understanding of the mechanism of synchronization.
The present analysis can also be applicable to models in higher dimensions
such as that for dislocations in solids which includes local nearest
neighbour interactions \cite{cap}. Furthermore, the present approach can be extended to understand the underlying mechanism in the case of locally coupled Kuramoto models with time delay \cite{10} (or phase delay) introduced between the coupled oscillators. In addition, the mechanism of synchronization in LCKM for open and fixed boundaries can be studied in a similar manner to the present work as well as for the case of unidirectional LCKM. We also want to mention that the scaling law given by eq.~(5) has been found experimentally in a transition
to phase synchronization in $CO_2$ lasers \cite{prl} and in electronic
circuits \cite{Zhu2001,eps}. On the other hand one can not make a direct comparison between the mechanism of synchronization discussed here in LCKM and the scaling law that has been found in experiments since the physical systems are not necessarily the same.

\acknowledgments

HFE thanks both of the School of Physics, Bharathidasan University, Tiruchirappalli, India and Abus Salam ICTP, Trieste, Italy, for hospitality during a part of this work. The work of PM is supported in part by Department of Science and Technology, Government of India (Ref. No. SR/FTP/PS-79/2005), Conselho Nacional de Desenvolvimento Cient\'{\i}fico e Tecnol\'ogico (CNPq), Brazil, and the Third World Academy of Sciences (TWAS), Italy. FFF acknowledges CNPq for financial support.


\begin{thebibliography}{99}
\bibitem{1} A. T. Winfree, \emph{Geometry of Biological Time} (Springer, New York, 1990).

\bibitem{2} C. W. Wu, \emph{Synchronization in Coupled Chaotic Circuits and Systems} (World Scientific, 2002).

\bibitem{14} S. H. Strogatz, \emph{Sync: The Emerging Science of Spontaneous Order} (Hyperion, 2003).

\bibitem{5} C. M. Gray, P. Koenig, A. K. Engel, and W. Singer, Nature (London) {\bf 338}, 334 (1989).

\bibitem{6} K. Otsuka, \emph{Nonlinear Dynamics in Optical Complex Systems} (Kluwer, Dordrecht, 2000).

\bibitem{10} H. Haken, \emph{Brain Dynamics: Synchronization and Activity Patterns in Pulse-Coupled Neural Nets with Delays and Noise} (Springer, Berlin, 2007).

\bibitem{13} M. Golubitsky and E. Knobloch Eds., \emph{Bifurcation, Patterns and Symmetry}, Physica D, {\bf 143} (2000).

\bibitem{15} Y. Kuramoto, \emph{Chemical Oscillations, Waves and Turbulences} (Springer, Berlin, 1984).

\bibitem{7} G. Hu, Y. Zhang, H. A. Cerdeira, and S. Chen, Phys. Rev. Lett. {\bf 85}, 3377 (2000).

\bibitem{8} Y. Zhang, G. Hu, H. A. Cerdeira, S. Chen, T. Braun, and Y. Yao, Phys. Rev. {\bf E 63}, 026211 (2001).

\bibitem{9} Y. Zhang, G. Hu, and H. A. Cerdeira, Phys. Rev. {\bf E 64}, 037203 (2001).

\bibitem{is} I. A. Heisler, T. Braun, Y. Zhang, G. Hu, H. A. Cerdeira, Chaos {\bf 13}, 185 (2003).

\bibitem{11} P. A. Tass, \emph{Phase Resetting in Medicine and Biology} (Springer, Berlin, 1999).

\bibitem{17} J. A. Acebron, L. L. Bonilla, C. J. P. Vicente, F. Ritort and R. Spigler, Rev. Mod. Phys. {\bf 77}, 137 (2005).

\bibitem{arx} Y. Ma and K. Yoshikawa, ArXive:0809.1697V3, nlin, (2008).

\bibitem{wie1} Y. Braiman, T. A. Kennedy, K. Wiesenfeld and A. Khinik, Phys. Rev. {\bf A}, 52, 1500, (1995).

\bibitem{16} B. C. Daniels, S. T. M. Dissanayake and B. R. Trees, Phys. Rev. E {\bf 67}, 026216 (2003).

\bibitem{22} Z. Liu, Y.-C. Lai and F. C. Hoppensteadt, Phys. Rev. {\bf E 63}, 055201(R) (2001).

\bibitem{wie2} A. Khinik, Y. Braiman, V. Protopopescu, T. A. Kennedy and K. Wiesenfeld, Phys. Rev. A {\bf 62}, 063815, (2000).

\bibitem{wie3} D. Tsygankov and K. Wiesenfeld, Phys. Rev. {\bf E}, 73, 026222, (2006).

\bibitem{suz}  S. Manrubbia, A. Mikhailov and D. Zanette, \emph{Emergence of dynamical Order: Synchronization Phenomena in Complex Systems} (World Scientific, Singapore, 2004).

\bibitem{iop} J. Rogge and D. Aeyels, J. Phys. A. {\bf 37}, 11135 (2004).

\bibitem{Strogatz1988} S. H. Strogatz and R. E. Mirollo, Physica D. {\bf 31}, 143 (1988).

\bibitem{Zheng98} Z. Zheng, G. Hu and B. Hu, Phys. Rev. Lett., {\bf 81}, 81 (1998).

\bibitem{Zheng2000} Z. Zheng, B. Hu and G. Hu, Phys. Rev. E {\bf 62}, 402 (2000).

\bibitem{19} H. F. El-Nashar, A. S. Elgazzar and H. A. Cerdeira, Int. J. Bifurcation and Chaos {\bf 12}, 2945 (2002).

\bibitem{m}  H. F. El-Nashar, Y. Zhang, H. A. Cerdeira and F. Ibyinka A., Chaos {\bf 13}, 1216 (2003).

\bibitem{20} H. F. El-Nashar, Int. J. Bifurcation and Chaos {\bf 13}, 3473 (2003).

\bibitem{21} P. Muruganandam, F. F. Ferreira, H. F. El-Nashar and H. A. Cerdeira, Pramana J. - Phys. {\bf 70}, 1143 (2008).

\bibitem{stro} S. H. Strogatz, \emph{Nonlinear Dynamics and Chaos} (Persus Publishing, 2000).

\bibitem{cap} A. Carpio and L. L. Bonilla, Phys. Rev. Lett. {\bf 90}, 135502-1 (2003).

\bibitem{prl} S. Bocaletti, E. Allaria, R. Meucci and F. Arecchi, Phys. Rev. Lett. {\bf 89}, 194101-1 (2002).

\bibitem{Zhu2001} L. Zhu, A. Raghu, Y.-C. Lai,  Phys. Rev. Lett. {\bf 86}, 4017 (2001)

\bibitem{eps} G.-M. Kim, G.-S. Yim, J.-W. Ryu, Y.-J. Park and D.-U. Hwang, Europhys. Lett. {\bf 71}, 723 (2005).

\end{thebibliography}
\end{document}